# Ultrafast atomic transport in recrystallizing ultrafine grained Ni

Daria Prokoshkina[a], Leonid Klinger[b], Anna Moros[a], Gerhard Wilde[a], Eugen Rabkin[b], and Sergiy V. Divinski[a,*]

[a]*Institute of Materials Physics, University of Münster, Wilhelm-Klemm-Str. 10, 48149 Münster, Germany*
[b]*Department of Materials Science and Engineering, Technion−Israel Institute of Technology, 32000 Haifa, Israel*

[*] Corresponding author: divin@uni-muenster.de

**Abstract**
We studied tracer self-diffusion in ultrafine grained Ni prepared by high pressure torsion. Two Ni materials of low (99.6 wt. %) and high (99.99 wt. %) purity levels were investigated. While the ultrafine grained structure of less pure Ni remained stable during diffusion annealing, recrystallization and subsequent grain growth occurred in high purity Ni at the same annealing conditions. Nevertheless, qualitatively similar ultrafast diffusion rates were measured in the samples of both purity levels. In high purity Ni, the kinetics of recrystallization was found to deviate strongly from the predictions of the Johnson-Mehl-Avrami-Kolmogorov theory. Moreover, the ultrafast diffusion paths withstood the recrystallization process. A model which accounts for solute redistribution in front of the moving boundary is suggested. Retaining of deformation-induced ultrafast diffusion paths in recrystallized Ni is explained by a specific mechanism of enhanced stability of the residual ultrafine grained fraction against recrystallization.

**Keywords**: high pressure torsion (HPT); nickel; diffusion; recrystallization; solute segregation

## 1. Introduction

Extensive investigations of ultra-fine grained (UFG) materials obtained by severe plastic deformation (SPD) during the last decades revealed many attractive kinetic properties of these materials related to ultrafast diffusion rates in the as-processed state [1-4]. The ultrafast diffusion was measured in UFG Ni [5], Cu-Zr [6] and Cu-Pb alloys [7], pure Cu [8], Ni [9] and Ti [10] prepared by Equal Channel Angular Pressing (ECAP). Simultaneously, it was found that SPD processing is often accompanied by the formation of a pronounced hierarchy of kinetic properties of the internal interfaces in UFG materials [6,11]. The idea of a "hierarchic microstructure" with agglomerates of small grains with relatively "slow" intra-agglomerate boundaries forming clusters with kinetically different ("fast") inter-agglomerate interfaces was originally suggested for clustered (agglomerated) nanocrystalline materials obtained through powder processing [12, 13]. Detailed measurements of the atomic diffusion by the radiotracer technique in SPD materials revealed characteristic concentration profiles typically for hierarchically organized materials. Such profiles are composed of a near-surface, "slow" branch (associated in UFG materials with general high-angle grain boundaries which are identical to the boundaries in well-annealed coarse-grained polycrystals) and a deeper branch which corresponds to "high-diffusivity" ("fast") paths. An extensive discussion of such paths in terms of existing levels of the hierarchy in SPD materials is given in [14].

In Part I of the present paper [15], the existence of ultrafast diffusion paths was established in UFG Ni of 99.99 wt.% (4N) and 99.6 wt.% (2N6) purity levels produced via high-pressure torsion (HPT). While the ultrafast diffusion paths in 2N6 Ni after HPT processing are qualitatively similar to their counterparts in ECAP-strained 2N6 Ni [9], the existence of these ultrafast paths in 4N Ni is very surprising in view of the complete recrystallization of this material during the corresponding diffusion annealing treatment. How do the ultrafast diffusion paths withstand the recrystallization and grain growth? What is the nature and mechanism of the unusual stability of these paths? These are the main subjects of the present study.

## 2. Experimental details
Materials and details of the deformation procedure were described in Part I. After high pressure torsion (HPT) at room temperature under a pressure of 2 GPa and applying 5 rotations, the average grain sizes of about 100 and 200 nm in 2N6 and 4N Ni, respectively, were determined.

*The microstructure* is characterized with a FEI Nova NanoSEM 230 scanning electron microscope (SEM) equipped with electron back scatter diffraction (EBSD) attachment, which was operated at 18 kV. The preparation of the sample surface was identical to that for the focused ion beam (FIB) analysis in Ref. [15]. For investigating the defect structure of deformed samples, a Zeiss Libra 200 FE field-emission transmission electron microscope (TEM) equipped with an in-column Omega energy filter was employed. The TEM samples of 3 mm in diameter were cut from the HPT processed disc at the half-radius distance from its center employing spark erosion. The samples were thinned to 60-90 μm thickness and finally electropolished using a solution of 17 % perchloric acid in ethanol at 10.5 V and a temperature of –21 °C.

*Differential scanning calorimetric* (DSC) measurements were performed using a Diamond DSC device (PerkinElmer, USA). Ni samples (at a half of the diameter of HPT processed samples) were cut into disks of 4 mm in diameter by spark erosion, grinded and cleaned in an ultrasonic bath. Heating was carried out in Ar atmosphere in a temperature range from 303 K to 723 K using the heating rate of 10 K/min. Three identical runs were done for each sample and it was proven that the second and third runs provide nearly identical results. The irreversible heat release was determined by subtracting the third-run signal from the first one.

*The radiotracer method* was used to investigate Ni self-diffusion. For a detailed description see Part I of the present paper [15].
The penetration profiles of self-diffusion in Ni deformed by HPT for Ni of both purities are shown in Fig. 1. For a better visualization they are plotted in coordinates of the relative specific activity *vs.* the penetration depth squared.
The striking feature of the penetration profiles measured for GB self-diffusion in UFG Ni of 4N purity is that they reveal systematically two distinct branches, a near-surface one at shallow depths below 4-6 μm; and a significantly deeper one. It is important that the first branch is nearly absent in the case of less pure 2N6 Ni (where it is almost exclusively caused by grinding-in effects) and it reveals a remarkable kink in the case of 4N Ni, which was discussed in Part I [15]. The present Part II is focused on the deep branches of the penetration profiles, which are related to the ultrafast atomic transport and, in fact, are observed in both materials of the two basically different purity levels.
The effective diffusivities, $D^f$, corresponding to the deep branches of the concentration profiles (the solid lines in Fig. 1) were determined employing the *erfc*-solution of the GB diffusion problem, i.e. assuming constant-source initial conditions,

$$\bar{c} = c_0 \, erfc\left(\frac{y}{2\sqrt{D^f t}}\right), \tag{1}$$

and they are listed in Table 1. In Eq. (1), $\bar{c}$ is the layer concentration which is proportional to the measured tracer activity, *y* is the penetration depth, and *t* is the diffusion time.

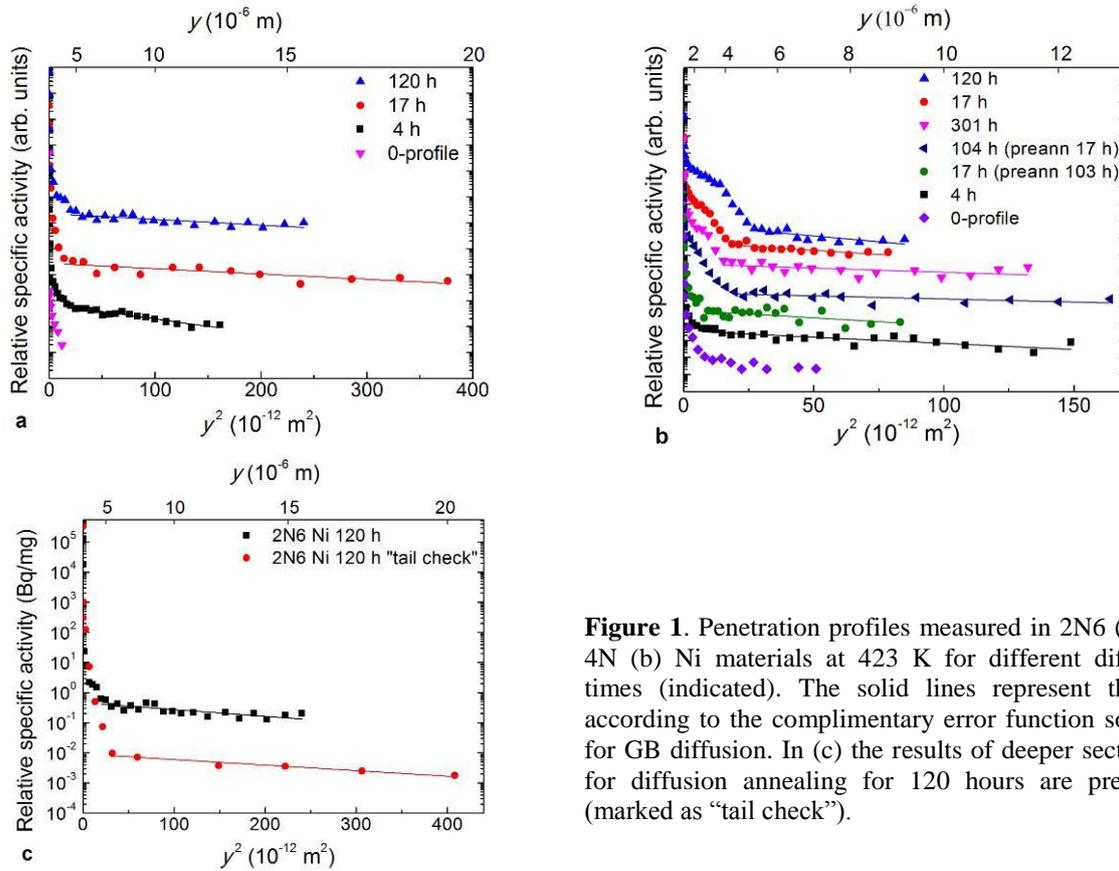

**Figure 1**. Penetration profiles measured in 2N6 (a) and 4N (b) Ni materials at 423 K for different diffusion times (indicated). The solid lines represent the fits according to the complimentary error function solution for GB diffusion. In (c) the results of deeper sectioning for diffusion annealing for 120 hours are presented (marked as "tail check").

## 3. Results and Discussion

The ultrahigh rates of diffusion could hypothetically be promoted by different mechanisms. One possible path is represented by so-called interconnected (percolating) porosity, when in addition to isolated voids (closed porosity was, e.g., reported for ECAP-ed Ti [16]), channels of voids penetrating deep into the material are observed. Such channels act as free surface with correspondingly high diffusivity. This type of porosity associated with SPD was observed in Cu [8] and Cu-1wt.% Pb alloy [7] produced by ECAP, but was absent in Ni ECAP-ed at room temperature [9], and in Ti severely deformed at 573 K [10]. However, in all these materials, the ultra-fast diffusion paths were observed.

### 3.1 Porosity channels as a reason for the ultrafast diffusion paths

In order to check the presence of open porosity in the HPT-deformed samples, a special experiment using the radioactive $^{65}$Zn isotope was performed. The reason of such choice for the tracer is the high energy of γ-quanta emitted by $^{65}$Zn (1.12 MeV), so that the presence of radioactive nuclei in the sample bulk could easily be detected due to the low self-absorption in the material. Note that the low-energy (70 keV) β-quanta emitted by $^{63}$Ni are strongly absorbed within the sample and the radioactivity of a near-surface layer could only be detected in this case. The $^{65}$Zn tracer was deposited by dropping 7 µl of liquid tracer solution on the polished specimen's surface and the solution was left for drying. Subsequently, a layer of about 10 µm in thickness was removed by grinding (the thickness was determined from the mass change due to grinding), the sample was carefully cleaned and the residual radioactivity was measured by an intrinsic Ge γ-detector of an extremely low background. A thickness of 10 μm was chosen to eliminate safely any possible artifacts due to zinc surface diffusion in nickel at room temperature. In the previous cases, when the presence of open porosity was detected, penetration depths of hundreds of micrometers were found. In the present case, no signal from the $^{65}$Zn tracer has been detected over the background. This observation indicates the absence of open (percolating) porosity in 4N Ni directly after room-temperature HPT deformation. The present result is consistent with that for ECAP Ni deformed at room temperature [9].

Table 1 – GB self-diffusion coefficients, $D^f$, determined for the ultrafast paths for UFG Ni of the two purity levels. Two diffusion experiments were performed after pre-annealing of the HPT Ni at 423 K for the indicated temperatures in order to initiate recrystallization and grain growth. Typical uncertainties of the measured $D^f$ values are below 15 %.

| Annealing time (h) | $D^f$ (m$^2$/s) | |
|---|---|---|
| | 2N6 Ni | 4N Ni |
| 4 | 1.8·10$^{-15}$ | 1.7·10$^{-15}$ |
| 17 | 1.7·10$^{-15}$ | 2.5·10$^{-16}$** |
| 17* | | 2.7·10$^{-15}$ |
| 17 (+103 h pre-anneal) | | 3.1·10$^{-16}$ |
| 104 (+17 h pre-anneal) | | 1.3·10$^{-16}$ |
| 120 | 2.4·10$^{-16}$ | 2.3·10$^{-17}$** |
| 120* | 1.9·10$^{-16}$ | 5.7·10$^{-16}$ |
| 301 | | 2.1·10$^{-17}$ |
| 301* | | 4.0·10$^{-17}$ |

*Diffusion experiments were repeated with sectioning to maximum depths in order to verify the values of the corresponding diffusion coefficients.
**On these samples sectioning was not completed.

## 3.2 Relaxation of the ultrafast diffusion paths

The previous experiments excluded pores as the ultrafast diffusion paths. Thus, the internal interfaces, which are probably in a specific deformation-modified (often referred to as "non-equilibrium" [17]) state introduced by SPD, exhibit accelerated GB diffusion [14].

A model of such "non-equilibrium" GBs was put forward by Nazarov et al. [17]. Using a disclination-based concept of such a state, the corresponding relaxation time, $\tau$, of the array of extrinsic GB dislocations can be estimated as [18]

$$\tau = \frac{d^3 kT}{A\delta D_{gb} G \Omega_a}. \qquad (2)$$

Here $d$ is the grain size; $G$ is the shear modulus; $\Omega_a$ is the atomic volume of Ni; $A$ is a geometrical factor of the model; $\delta$ is the GB width; and $kT$ has its usual meaning. The key problem of this expression is represented by the grain boundary diffusion coefficient, $D_{gb}$, which has to be employed in estimates. In Ref. [9] it was shown that it is the diffusion coefficient of *relaxed* high-angle GBs, as they are present in well-annealed high purity polycrystalline Ni, which has to be used in order to obtain data that fit to the experimental observations. Using further the value of $A=200$ [9], the relaxation times are estimated to be equal to 7 and 56 hours in 2N6 and 4N Ni, respectively. This relatively large difference between the two materials stems from the strong dependence of $\tau$ on the grain size, Eq. (2).

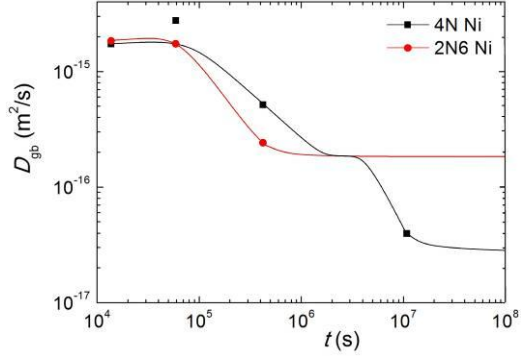

**Figure 2**. Diffusion coefficients as a function of time (anneals at 423 K). The curves are drawn only as a guide for the eye and correspond to the present understanding of a two-stage relaxation of the deformation-modified state of high-angle GBs in HPT Ni, see text.

These estimates are in acceptable agreement with the present experimental diffusion data – the diffusion coefficients are nearly equal for 4 and 17 hour annealing treatments, and decrease by an order of magnitude after diffusion annealing for more than 120 hours, Fig. 2. This decrease may be explained by the relaxation of the deformation-modified interfaces in HPT Ni. However, in line with our previous observation in ECAP-processed 2N6 Ni [9], these values do not approach those for relaxed high-angle GBs (which would be about $10^{-21}$ m$^2$/s or even less, depending on the segregation level of residual impurities), but represent a modified, metastable state of the "non-equilibrium" GBs. The existence of such metastable non-equilibrium GB states was predicted in atomistic computer simulations of Rittner and Seidman [19]. We propose that because of the relatively low homologous temperature of the diffusion experiment (about $0.24T_m$, where $T_m$ is the melting point of Ni) the nucleation rate of truly relaxed GB "phase" may be negligibly small and the GBs can preserve their metastable configuration for a long time.

The dependence of the derived diffusion coefficients on the diffusion time could be explicitly visualized by re-plotting the concentration profiles against the reduced depth, $y^2/t$. In the case of a constant diffusivity, the penetration profiles have to follow a "master" plot, see Eq. (1), and since at larger depth the *erfc*- and Gaussian-solutions are similar, one should get a set of almost parallel lines.

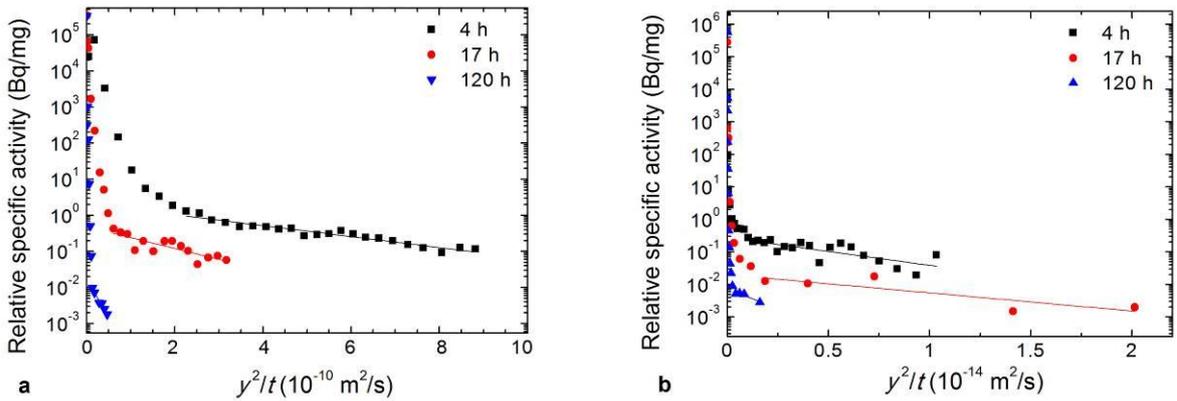

**Figure 3**. Penetration profiles in 2N6 (a) and 4N (b) Ni after anneals at 423 K for the indicated times, plotted against the reduced depth $y^2/t$.

Figure 3 shows the penetration profiles in the reduced coordinates, and for times of 120 hours a characteristic change of the corresponding slopes at large depths can clearly be seen.

## 3.3 Kinetics of recrystallization of 4N Ni

In Part I it was shown that during isothermal annealing at 423 K of 4N Ni, the heat flow after 130 hours of annealing becomes very small (the obtained calorimetric signal was less than 1 μW) and close to the thermal noise level, indicating that the recrystallization processes are complete (see Fig. 4b in [15]). The amount of unrecrystallized UFG nickel after 130 hours of annealing is almost zero when calculated using the corresponding JMAK expression, Eq. (2) in [15], although microstructure examination indicated the existence of a residual UFG fraction even after such long annealing.

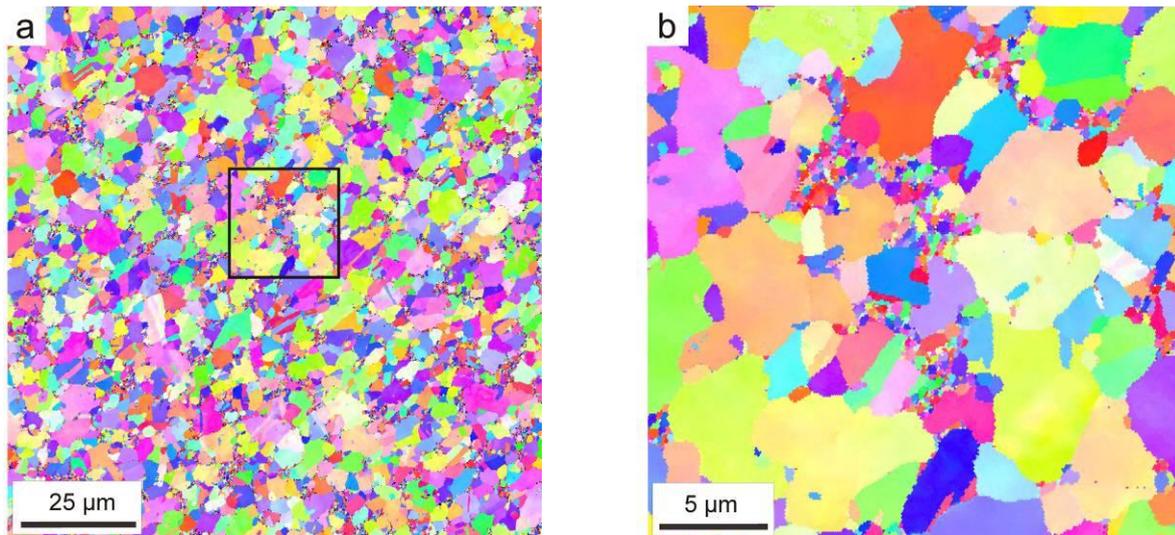

**Figure 4**. EBSD images of (a) 4N HPT Ni annealed at 423 K during 2 weeks and (b) the enlarged area marked by the black square in (a).

In order to check whether recrystallization was indeed fully accomplished after 130 hours of annealing at 423 K or the signal from the recrystallization processes is simply too weak to be detected by calorimetric measurements, EBSD investigations were performed. A homogeneous coarse-grained microstructure is observed with a small amount of islands of residual ultrafine unrecrystallized grains, Fig. 4. The average volume fraction of the unrecrystallized fraction, counting the grains with the sizes between 100 to 700 nm, is about 10 %, Fig. 4a. Even after extended annealing for 3 weeks, the residual UFG regions still account for up to 6 % of the total volume, Fig. 4b. Thus, recrystallization continues, but with a significantly slower rate compared to that predicted by the JMAK equation. Similar deviations from the JMAK equation for long annealing times were observed in several previous studies (see, for example, [20]).

The UFG fraction of the microstructure is not randomly distributed within the material, but is concentrated in elongated clusters. In fact, these clusters look like chains of small grains, which extend through the material, Fig. 4a.

The grain size distribution, shown in Fig. 5, reveals that the grain size is $(200 \pm 100)$ nm in the as-prepared state, while a bimodal distribution is obvious in the annealed material. There is a fraction of relatively small grains, which form the non-recrystallized areas, with grain sizes well below 1 μm. The other microstructure component is represented by relatively large recrystallized grains of about $(5 \pm 3)$ μm in diameter. Note that the smallest grains in the annealed microstructure could not be resolved in Fig. 5 due to the chosen measurement step size of 100 nm which is a compromise between the requirement to obtain a good statistic for both small and large grains in recrystallized microstructure.

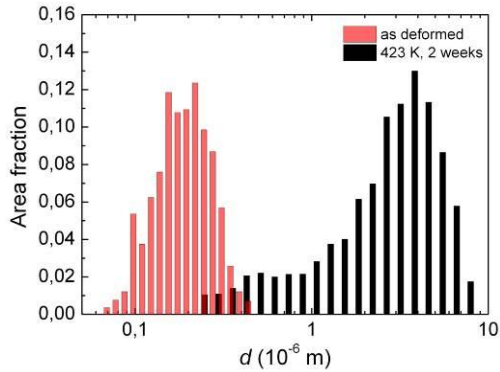

**Figure 5**. The grain size distribution (area fraction) in 4N Ni in the as-prepared state and after an additional annealing treatment at 423 K for 2 weeks.

*3.2 Calorimetric measurements*

The kinetics of defect recovery during annealing treatments was investigated by DSC measurements. The examples of DSC curves measured for HPT Ni of the two purity levels in the temperature interval of 350–575 K (i.e. below the Curie temperature) are presented in Fig. 6 for the heating rate of 10 K/min.

The DSC curves for HPT-processed 2N6 Ni are very similar to those measured previously in this materials after ECAP deformation [9] and correspond to three basic processes − vacancy annihilation around 400 K, annihilation of vacancy–impurity complexes and vacancy clusters between 450 and 500 K, and dislocation annihilation and re-arrangements between 500 and 550 K, Fig. 6.

The DSC signal in HPT-deformed 4N Ni is different and reveals a dominant contribution of recrystallization processes between 500 and 550 K in addition to a moderate signal from vacancy annihilation around 400 K. This behavior is similar to that observed in HPT 4N Ni by Setman et al. [21].

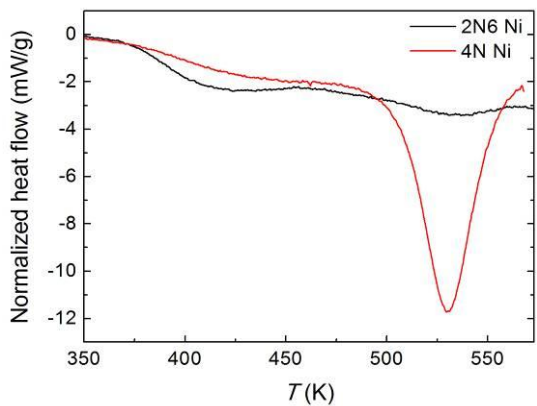

**Figure 6**. Basis line-corrected DSC curves obtained at 10 K/min linear heating of 4N and 2N6 HPT nickel.

The effect of defect recovery and recrystallization on the structure of grain boundaries, which are presumably the pathways for short-circuit diffusion in 2N6 and 4N Ni, was further examined by TEM.

*3.4 Microstructure and grain boundary structure as revealed by TEM*

The microstructure of the HPT samples was examined by TEM in the as-prepared state and after heating to pre-selected temperatures with a constant heating rate (used in the DSC analysis) and subsequent rapid quenching. As an example, the TEM images of 4N and 2N6 Ni in the as-prepared state and after heating up to 573 K and subsequent quenching are shown in Fig. 7.

The samples of UFG Ni of both purity levels reveal generally similar microstructures with two types of grain boundaries, namely relatively straight and flat boundaries and interfaces with strong serrations, Fig. 7a, c. The latter have been suggested to be responsible for accelerated GB diffusion in SPD materials, see e.g. [9, 14]. After the heat treatment, the difference between the Ni samples of the two purity levels is evident, Fig. 7d, g. Recrystallization is completed in 4N Ni, grains are almost defect-free, the grain size has grown by a factor of ten to twenty, and almost perfect straight grain boundaries are exclusively observed. In the case of 2N6 Ni, the GB structure appears to be more relaxed than in the as-prepared state, the interfaces with a serrated contrast are rarely seen, the dislocation density is significantly reduced, but still no grain growth occurred. A partial relaxation of GB structure in the 2N6 Ni is probably caused by mobile vacancies (see Fig. 6).

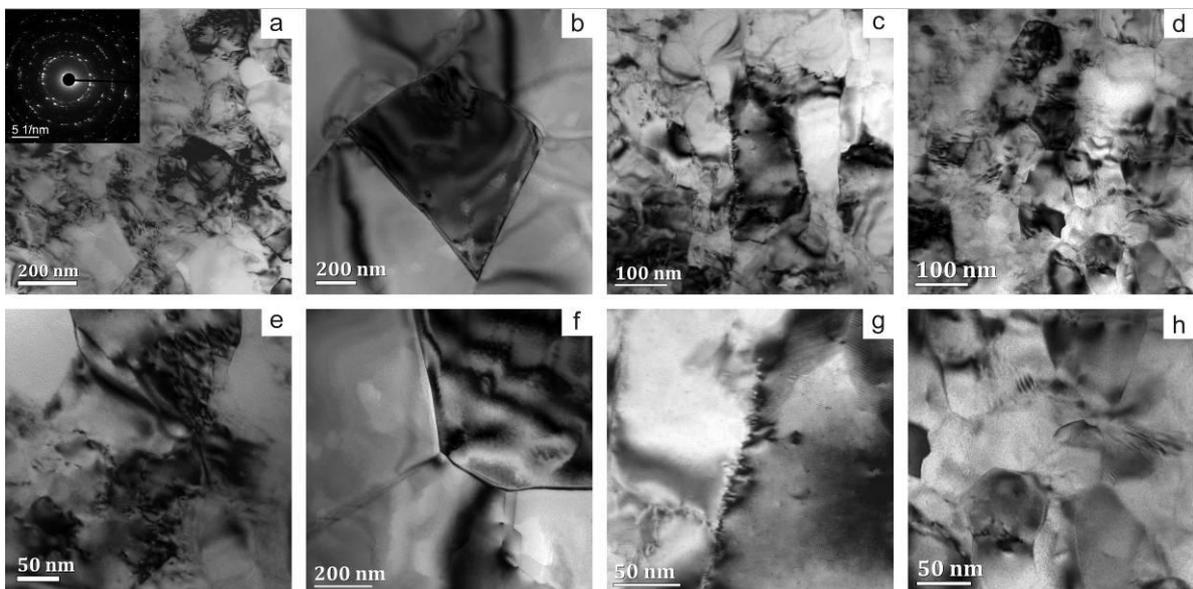

**Figure 7**. TEM micrographs of Ni samples of 4N (a, b) and 2N6 (c, d) purity directly after HPT deformation (a, c), and after linear heating to 573 K and subsequent quenching (b, d). Figures (e-h) represent enlarged areas of the upper micrographs.

Taking into account the evolution of the microstructure and the estimates of the relaxation time, the measured diffusion penetration profiles can be treated as follows: the first branch of the concentration profile is affected by recrystallization and the moving recrystallization front (see Part I [15]); while the second (deep) branch characterizes the diffusion along "fast" GBs.

The second branch of the profiles is different in 2N6 and 4N Ni and evolves in a characteristic fashion with annealing time. In 2N6 Ni, the main process occurring during the first 120 h of annealing is the relaxation of GBs, during which the GB self-diffusion coefficient decreases by one order of magnitude (from $10^{-15}$ to $10^{-16}$ m$^2$/s). Afterwards the microstructure remains practically unchanged and the grain boundary diffusion rate does not evolve further for accessible diffusion times.

In 4N Ni, the evolution of the GB diffusivity occurs in two steps, i.e. additionally to the GB relaxation similar to that in 2N6 Ni at intermediate annealing times, a redistribution of impurities due to the vacancy flux and/or recrystallization takes place. We suppose that these impurities are then absorbed by the GBs and further slow down the diffusing flux, Fig. 3. At the same time these redistributed impurities block the recrystallization front and cause deviations from the JMAK recrystallization kinetics, see below. This increase in impurity content at the GBs might lead to a further decrease of the GB self-diffusion coefficient by one order of magnitude (from $10^{-16}$ to $10^{-17}$ m$^2$/s, Table 1).

We emphasize that the grain size plays a decisive role concerning the effect of impurities on GB diffusion. In 2N6 Ni the average grain size is significantly smaller than in 4N Ni (by a factor of 2 to 3), which makes the GBs in 2N6 Ni almost impurity-free (concerning the explanation of the effective purification of GBs with decreasing grain size, see Ref. [22]). Since the grains in 2N6 Ni do not grow during annealing, the GBs remain impurity-free. At the same time the recrystallization and grain growth in 4N Ni causes the redistribution of impurities which, in turn, affects the kinetics of recrystallization and the GB diffusivity. In what follows, we will present a quantitative model of these phenomena.

### 3.5. Model

The re-distribution of impurities at the GBs has been considered in the past as a possible reason for stagnation [23] or slow-down [24] of grain growth in nanocrystalline materials. Yet we demonstrated in the present work that HPT-processed 4N Ni undergoes recrystallization during annealing at 423 K, with little changes occurring in the remaining UFG structural component. The re-distribution of impurities during recrystallization is expected to be very different from that occurring during the normal grain growth. A characteristic stagnation of recrystallization in the long-time limit observed in this work is most probably related to the impurity redistribution in front of the moving boundary which separates the residual UFG component and the recrystallized grains. The retardation becomes most obvious when two recrystallization fronts approach each other, because of the accumulation of impurities in the UFG area between the fronts.

This type of enhanced stability of the UFG component against the consumption by the approaching recrystallization front is different from similar behavior which was observed at the initial stages of recrystallization in ECAP-processed 2N6 Ni [25]. In the latter case the existence of specific rotational defects in the grain body, which correspond to lattice bending with the rotational axis perpendicular to the high-angle grain boundary, induces an enhanced stability of such microstructure elements against recrystallization and grain growth via pinning by orientation gradients [25], as observed also for severely deformed nanocrystalline Pd [26].

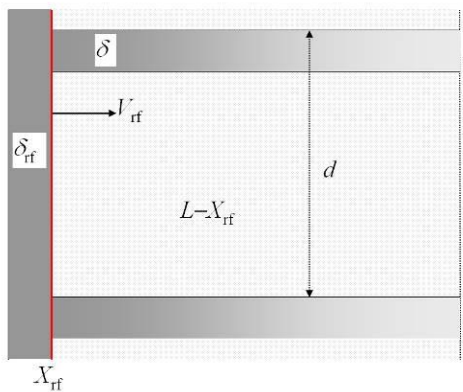

**Figure 8**. A scheme of the microstructure near a recrystallization front (in red) moving with the velocity $V_{rf}$. $d$ is the grain size; $\delta$ and $\delta_{rf}$ are the widths of grain boundary and recrystallization front, respectively; $L$ is the half distance between the approaching recrystallization fronts; and $X_{rf}$ is a current position of the left front. The darkness of the grey areas corresponds schematically to the local concentrations of the solute.

In order to describe the impact of GB diffusion of the impurity atoms on the normal motion of the grain boundaries during recrystallization, the following scheme (see Fig. 8) is considered: the moving GB (recrystallization front) accumulates all impurities of the bulk and of the GBs between the ultrafine grains in the UFG matrix to be consumed, and thus the concentration of the impurities in the GB increases. For simplicity, we will assume that $\delta_{rf} \approx \delta$, where $\delta_{rf}$ and $\delta$ are the thicknesses of the recrystallization front and of the GBs in the UFG matrix. The increasing GB concentration of impurities with a strong segregation tendency to the GB slows down the GB motion because these impurities are dragged by the migrating GB. A full description of a one-dimensional solute drag problem was given by Cahn [27], yet his treatment employs solution thermodynamics in the vicinity of moving GB. The use of equilibrium thermodynamics is questionable in the present case of annealing at 423 K, the temperature at which the atoms in the bulk of Ni do not move. Therefore, we employ here a simpler approach similar to the one of Lücke and Detert [28]. Firstly, we will assume that the driving force for recrystallization, $P$, remains constant for the whole annealing duration. Then, the recrystallization front velocity in the impurity-free material, $V_{rf}^0$, is

$$V_{rf}^0 = M_0 P \qquad (3)$$

where $M_0$ is the recrystallization front mobility in pure material. The impurity atoms migrating together with the recrystallization front exercise the drag force on this front, $f_i$, which is a ratio of front velocity and diffusion mobility of impurity atom:

$$f_i = \frac{kT}{D_i} V_{rf} \qquad (4)$$

where $V_{rf}$ and $D_i$ are the velocity of impurity-laden front and the diffusion coefficient of impurity atoms *normal* to the GB, respectively. Thus, the net driving force acting on the moving recrystallization front is reduced by the drag force due to impurity atoms dragged along:

$$V_{rf} = M_0 \left( P - \frac{n_b \delta k T C_{rf}}{D_i} V_{rf} \right) \qquad (5)$$

where $n_b$ and $C_{rf}$ are the number of Ni atoms per unit volume of the GB phase and the molar fraction of impurities in the moving GB, respectively. Here, we assumed that the impurity concentration in the moving GB is sufficiently small so that the intrinsic GB mobility and diffusion coefficient of the impurity atoms do not change [29]. Combining Eqs. (3)–(5) yields the following expression for the front velocity:

$$V_{rf} \equiv \frac{dX_{rf}}{dt} = \frac{V_{rf}^0}{1 + \alpha \cdot C_{rf}}, \text{ with } \alpha = \frac{n_b \delta k T M_0}{D_i} \qquad (6)$$

The estimation of the value of parameter α is difficult because of uncertainties in the values of $M_0$ and $D_i$. Assuming that most of the driving force for recrystallization stems from the consumption of GBs in the UFG matrix by the recrystallization front, and estimating $D_i$ by the self-diffusion coefficient *along* the relaxed GBs in coarse grain Ni yields α ≈ 100, which means that only 1 at. % of impurities in the migrating GB (well below the detection limit of analytical TEM) decreases its migration rate by a factor of two.

The accumulation of impurities in the migrating recrystallization front creates a driving force for the forward GB diffusion along the GBs of the untransformed UFG matrix:

$$\frac{\partial C}{\partial t} = D \frac{\partial^2 C}{\partial x^2}, \; C\big|_{x=X_{rf}} = C_{rf}, \; \frac{\partial C}{\partial x}\bigg|_{x=L} = 0 \qquad (7)$$

$$\frac{dC_{rf}}{dt} d\delta = V_{rf} \left[ C_{rf} \delta + C_0 (d - \delta) \right] + D\delta \left( \frac{\partial C}{\partial x} \right)_{x=X_{rf}} \qquad (8)$$

where $C_0$, $C$ and $D$ are the bulk concentration of the impurities, the concentration of impurities in the GBs of the untransformed UFG matrix, and the diffusion coefficient of impurities along these GBs, respectively. Here it is assumed that the diffusion coefficient of the moving GB, $D_{mb}$, is sufficiently large, $\frac{D_{mb}}{d \cdot V_{rf}} \gg 1$, and impurities from the GBs in the UFG matrix are immediately redistributed along the moving GB. This assumption follows from the fact that the moving GB accumulates defects from the UFG matrix (dislocations), acquires extra free volume and increases its diffusivity (see Part I [15]). A small value of $D_{mb}$ will force the condition (a) described below, even for large values of $D$.

We suppose that initially the system is uniform

$$C|_{t=0} = C_{rf}|_{t=0} = C_0 \qquad (9)$$

We can consider two limiting cases.

a) $D/LV_{rf}(C_0) \ll 1$. In this case, an additional amount of the impurities accumulated outside of the moving GB can be neglected and the GB concentration can be calculated from the mass balance $\delta(C_{rf} - C_0) = X_{rf}C_0$ or

$$V_{rf} = \frac{V_{rf}^0}{1 + \alpha C_0 (1 + X_{rf}/\delta)} \qquad (10)$$

b) $D/LV_{rf}(C_0) \gg 1$. In this case, the gradient of the concentration along the forward GB can be neglected and the GB concentration can be calculated from the mass balance $(C_{rf} - C_0)\delta(d + L - X_{rf}) = C_0 X_{rf} d$ or

$$V_{rf} = \frac{V_{rf}^0}{1 + \alpha C_0 (1 + X_{rf} d/(d + L - X_{rf})\delta)} \qquad (11)$$

Figure 9 shows the results of the numerical solution of Eqs. (6)-(9) in the general case. The calculations were performed for the following values of relevant parameters: $V_{rf}^0 = 10^{-11}$ m/s, $D = 2 \cdot 10^{-17}$ m$^2$/s (see Table 1), $\delta$=0.5 nm, $d = 200$ nm, and $L = 2.5 \cdot 10^{-6}$ m. One can see in Fig. 9 that once two recrystallization fronts approach each other, their migration rate significantly slows down due to the enhanced impurity drag. This enhanced drag is a direct result of overlapping GB diffusion fields developing ahead of two approaching fronts. The role of this overlap is underlined by dashed lines in Fig. 9 showing the migration kinetics of individual recrystallization fronts. Though some moderate slowdown is also observed for individual fronts, it is much weaker than in the case of two converging fronts because the impurities at a single front are free to diffuse away in the direction of the front migration.

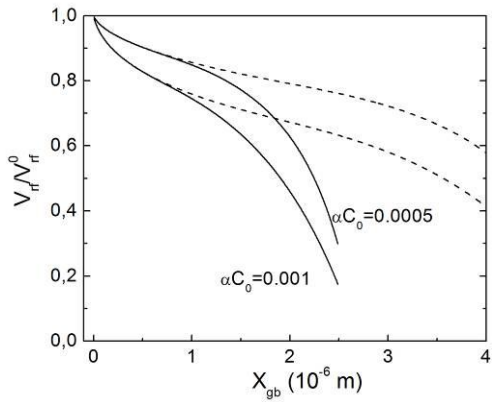

**Figure 9**. Characteristic slowdown of the migration rate, $V_{rf}$, of two converging recrystallization fronts as a result of impurity redistribution. The dashed lines show the migration kinetics of individual recrystallization fronts. $V_{rf}^0$ is the migration rate of the recrystallization front in absence of solutes; $C_0$ is the bulk concentration of solute and $\alpha$ is determined via Eq. (6).

The slowdown of front migration shown in Fig. 9 is consistent with the deviations from the JMAK equation for long annealing times described in Section 3.3, and with the elongated morphology of remaining UFG clusters observed in the EBSD images (see Fig. 4). In fact, these remaining percolating clusters of the UFG microstructural component may contain partially relaxed, metastable GBs ensuring ultrafast diffusivity of the recrystallized material. High impurity concentrations in these remaining GBs ensure an additional decrease of their diffusivity (see Fig. 2).

It should be noted that the increased concentration of impurities in the remnants of the UFG matrix predicted by our model may cause a corresponding decrease of the GB energy and, consequently, of the driving force for recrystallization [30]. However, a quantitative analysis of this phenomenon is hardly possible in the present case because the Gibbs segregation equation is not applicable (the GBs are not equilibrated with the bulk because of the lack of bulk mobility).

**Summary**
Ultrafast diffusion rates were measured in HPT-processed UFG Ni materials of two purity levels, 99.6 and 99.99 wt.%, which exhibited very different microstructure evolutions during diffusion annealing treatments at 423 K. Whereas the UFG microstructure in the 2N6 Ni was stable, recrystallization and subsequent grain growth occurred in 4N Ni. Surprisingly, ultrafast diffusion was observed in 4N Ni even after long annealing times and almost complete recrystallization, too. The studies of the microstructure of the 4N Ni samples after long annealing times revealed a significant retained fraction of the untransformed UFG matrix, well beyond the prediction of the JMAK model based on calorimetric measurements.

We proposed a model which takes into account the redistribution of impurities in front of the moving recrystallization front. The increase in impurity concentration between the two converging recrystallization fronts slows down their motion because of the impurity drag effect. Thus the remnants of the UFG matrix survive recrystallization in the form of narrow and elongated clusters separating recrystallized areas. The increase in impurity concentration in the remaining GBs of the UFG matrix causes an additional (in addition to the thermal relaxation) decrease of their ultrafast diffusivity. This additional relaxation of ultrafast diffusion paths was not observed in 2N6 Ni with a static microstructure, in which a moderate decrease of the ultrafast diffusion rates is satisfactory explained by the thermal relaxation of the deformation-modified state of grain boundaries.

**Acknowledgment**
This research was supported by a Grant from the GIF, the German-Israeli Foundation for Scientific Research and Development (Grant No. G-1037-38.10/2009) and partially by the Deutsche Forschungsgemeinschaft.